\documentclass[letterpaper]{article} 
\usepackage[submission]{aaai24}  
\usepackage{times}  
\usepackage{helvet}  
\usepackage{courier}  
\usepackage[hyphens]{url}  
\usepackage{graphicx} 
\usepackage{natbib}  
\usepackage{caption} 
\usepackage{algorithm}
\usepackage{algorithmic}
\usepackage{amssymb}
\usepackage{amsmath}
\usepackage{array}
\usepackage{pifont}
\newcommand{\cmark}{\ding{51}}%
\newcommand{\xmark}{\ding{55}}%

\usepackage{newfloat}
\usepackage{listings}
\DeclareCaptionStyle{ruled}{labelfont=normalfont,labelsep=colon,strut=off} 
\floatstyle{ruled}
\newfloat{listing}{tb}{lst}{}
\floatname{listing}{Listing}
\title{Oral-3Dv2: 3D Oral Reconstruction from Panoramic X-Ray Imaging with Implicit Neural Representation}

\author {
    Weinan Song\textsuperscript{\rm 1},
    Haoxin Zheng\textsuperscript{\rm 1},
    Dezhan Tu\textsuperscript{\rm 1},
    Chengwen Liang\textsuperscript{\rm 3},
    Lei He\textsuperscript{\rm 2, \rm 1}
}
\affiliations {
    \textsuperscript{\rm 1}University of California, Los Angeles\\
    \textsuperscript{\rm 2}Eastern Institute for Advanced Study, Ningbo, China\\
    \textsuperscript{\rm 3}Hangzhou Dental Hospital, Hangzhou, China
}

\usepackage{bibentry}

\begin{document}

\maketitle

\begin{abstract}
3D reconstruction of medical imaging from 2D images has become an increasingly interesting topic with the development of deep learning models in recent years. Previous studies in 3D reconstruction from limited X-ray images mainly rely on learning from paired 2D and 3D images, where the reconstruction quality relies on the scale and variation of collected data. This has brought significant challenges in the collection of training data, as only a tiny fraction of patients take two types of radiation examinations in the same period. Although simulation from higher-dimension images could solve this problem, the variance between real and simulated data could bring great uncertainty at the same time. In oral reconstruction, the situation becomes more challenging as only a single panoramic X-ray image is available, where models need to infer the curved shape by prior individual knowledge. To overcome these limitations, we propose Oral-3Dv2 to solve this cross-dimension translation problem in dental healthcare by learning solely on projection information, i.e., the projection image and trajectory of the X-ray tube. Our model learns to represent the 3D oral structure in an implicit way by mapping 2D coordinates into density values of voxels in the 3D space. To improve efficiency and effectiveness, we utilize a multi-head model that predicts a bunch of voxel values in 3D space simultaneously from a 2D coordinate in the axial plane and the dynamic sampling strategy to refine details of the density distribution in the reconstruction result. Extensive experiments in simulated and real data show that our model significantly outperforms existing state-of-the-art models without learning from paired images or prior individual knowledge. To the best of our knowledge, this is the first work of a non-adversarial-learning-based model in 3D radiology reconstruction from a single panoramic X-ray image.
\end{abstract}

\section{Introduction}
Radiological 3D reconstruction from limited 2D images has attracted increasing attention with the development of deep generative models in the past few years. Recent works like \cite{oral_3d, x2ct_gan, x_to_3d, 3d_leg} have shown the feasibility of 3D reconstruction from only one or two X-ray images, which provides an alternative solution to 3D imaging where only 2D imaging equipment is available. Due to the low radiation generated by 2D imaging equipment, these methods also bring a new choice in radiological examination for patients who are sensitive to radiation. For example, research in \cite{CT_radiation} shows that the X-ray imaging method could take as much as 200 less radiation than Cone Beam Computed Tomography (CBCT), a fast and low-radiation type of Computed Tomography (CT) and is widely used in dental radiology. Therefore, developing fast and accurate translation models could potentially bring great progress in medical imaging.

\begin{figure}[t]
    \centering
    \includegraphics[width=0.45\textwidth]{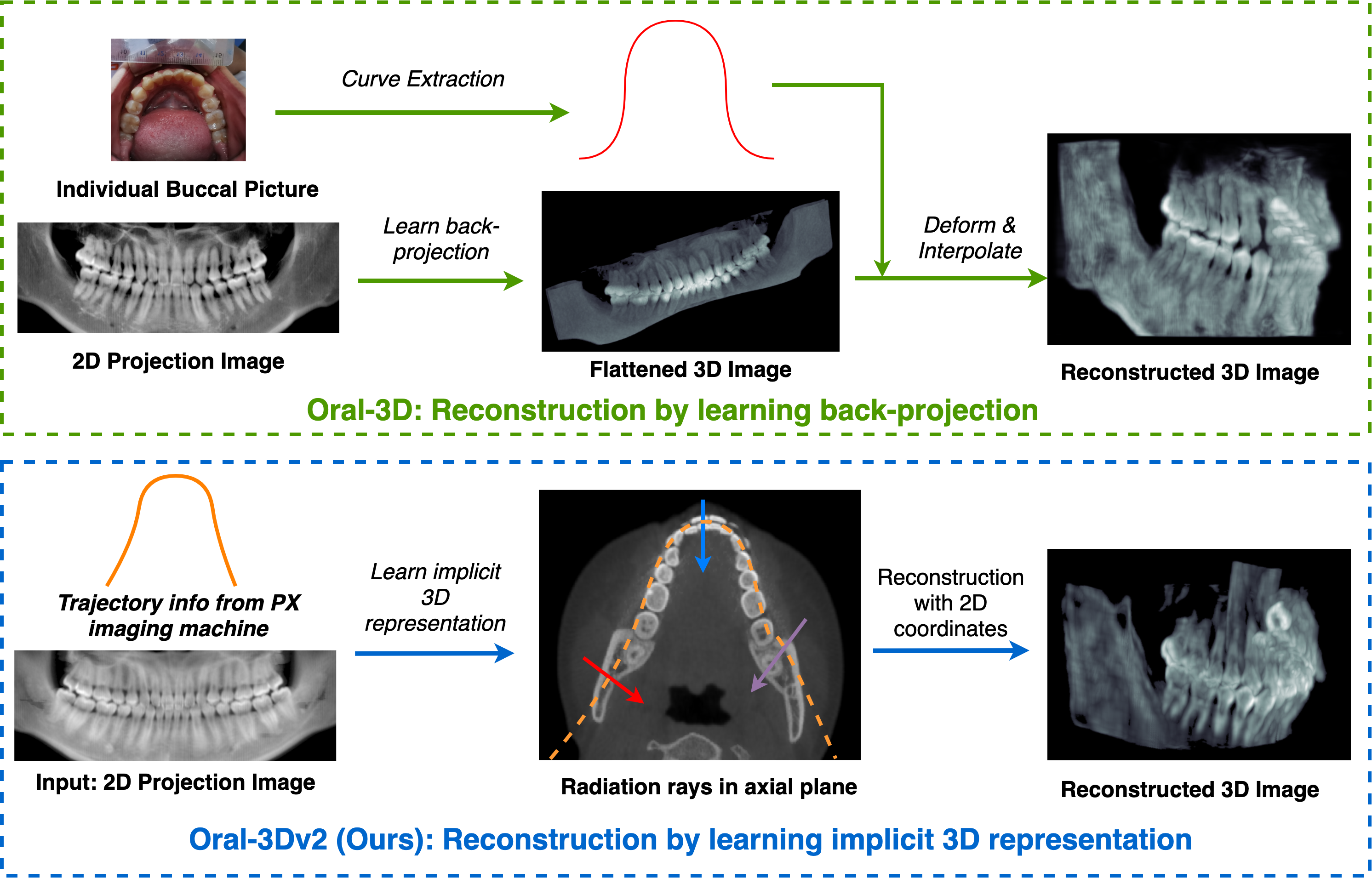}
    \caption{We compare our new model (blue) and Oral-3D (green) in this picture. Oral-3D first learns a back-projection model with paired images to generate a flattened 3D oral structure. Then it deforms the flattened image into a curved shape according to the individual dental arch shape acquired from the patient. In our model, we learn an implicit 3D representation of the oral structure only from the projection information, i.e., projection image and X-ray tube trajectory that is pre-defined by the equipment manufacturer and independent of individuality. After the model is well-trained, the 3D object is reconstructed by inferring the density distribution in 3D space from the implicit representation model and 2D coordinates. 
    }
    \label{fig:oral3d}
\end{figure}

\begin{figure*}[ht]
    \centering
    \includegraphics[width=0.9\textwidth]{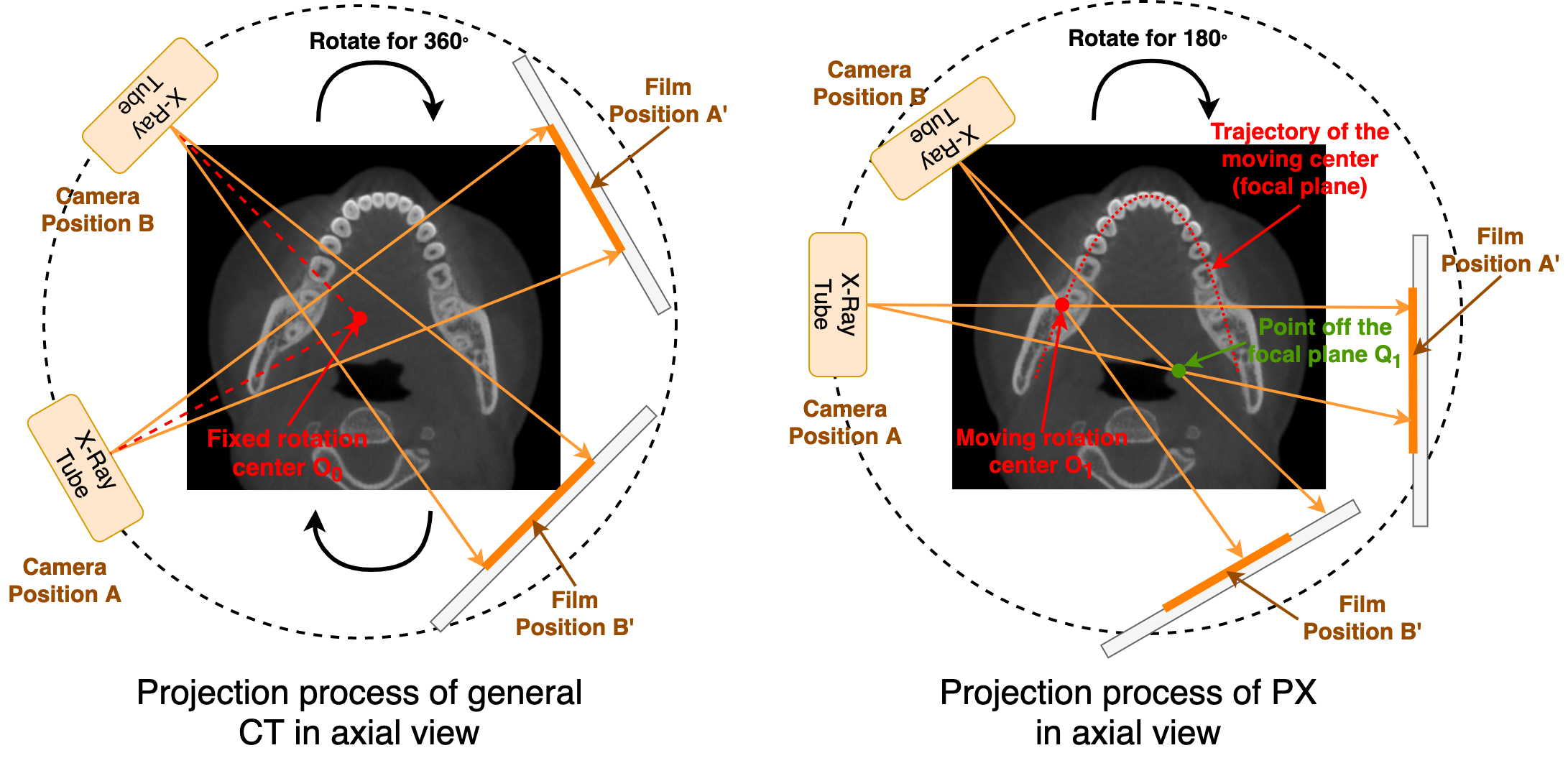}
    \caption{We show the comparison of imaging process of general CT (including CBCT) and PX in this picture. In CT, the X-ray tube and the film moves together around a fixed rotation center for $360$ degrees, where the film receives all X-rays sent from the tube. In PX imaging, the X-ray tube and the film rotates around a moving center, whose trajectory fits the curve of the mandible. Therefore, points that are around and away from the trajectory receive different levels of radiation during the imaging. For example, when the tube and the film moves from A to B in the right picture, the red point is projected twice while the green point is only projected once. This could make the image show more information of the imaging target at the red point over the green point.
    }
    \label{fig:imaging}
\end{figure*}

However, most of these cross-dimension translation models learn to explicitly generate a 3D image by auto-encoding and adversarial learning from paired X-ray images and CT scans. Consequently, the reconstruction quality is sensitive to the diversity and scale of training data. In dental imaging, restoring curved mandibular shapes brings additional challenges as only a single panoramic X-ray (PX) image is available. To solve this problem, recent studies like Oral-3D \cite{oral_3d} and X2Teeth \cite{x2_teeth} utilize individual prior knowledge when training the model, i.e., dental arch shape extracted from buccal images or instance annotations of teeth at the pixel level. Yet these complicated operations could bring conspicuous miss alignment during reconstruction, thus greatly hindering clinical applications in dental examinations. As a comparison, implicit representation models \cite{nerf, hash_nerf} provide a new solution in 3D reconstruction from 2D images. But these models rely on learning from abundant images viewed from various directions, which is hard to apply in radiology due to differences in imaging principles and inflexibility in imaging angles.
 
To address these limitations, we propose a new framework for 3D oral reconstruction from a single 2D panoramic X-ray (PX) image. Different from previous work like Oral-3D, which learns a back projection function to explicitly predict the reconstruction result by learning from paired images and prior knowledge of the individual dental arch shape, our model could learn 3D reconstruction simply from a single X-ray image with the projection settings from the imaging equipment. A comparison between Oral-3D and our method can be seen in Fig. \ref{fig:oral3d}, where only projection data is required during the reconstruction in our method.  

Unlike models in \cite{mednerf, xtransct} that utilize a single X-ray or two orthogonal X-ray images, our method could utilize the rich projection information during a panoramic scan with our advanced architecture. Specifically, we use a deep learning network to learn a mapping function between coordinates and density values of voxels in the 3D space, i.e., Hounsfield Unit (HU). To take advantage of the imaging process in panoramic imaging, we propose a multi-head model that outputs a bunch of voxel values at the same time given a 2D coordinate, which proves to be both efficient and effective over existing implicit representation models. Furthermore, to accommodate the imaging object in radiology, we utilize a dynamic sampling strategy to improve the reconstruction quality by acquiring points along radiation rays in random resolutions. Extensive experiments show that our model could significantly outperform state-of-the-art methods in 3D oral reconstruction both qualitatively and quantitatively. In conclusion, we summarize our contribution as follows:

\begin{itemize}
    \item Different from previous approaches in 3D oral reconstruction, such as Oral-3D\cite{oral_3d} and X2Teeth\cite{x2_teeth}, our model could achieve superior performance without training from any paired data, individual prior knowledge, or annotations.
     
    \item We propose an efficient implicit 3D representation model that maps a 2D coordinate into a bunch of 3D density values. This could reduce the computation complexity from $\mathcal{O}(N^3)$ to $\mathcal{O}(N^2)$ when reconstructing a $N \times N \times N$ object during both training and inference.
    
    \item We also propose a dynamic sampling strategy when sampling points from radiation rays with an adaptive projection method. This could encourage the model for higher reconstruction quality by learning a smooth density distribution in the 3D space .
\end{itemize}

\begin{table*}[hbt!]
    \centering
     \caption{Comparison of PX and CBCT in dental imaging}
\begin{tabular}{p{1.5cm}<{\centering}p{1.5cm}<{\centering}p{2.6cm}<{\centering}p{5cm}<{\centering}p{5.0cm}<{\centering}}
    \hline
    Method&Image&Rotation&Tomography Theory&Dental Applications\cr
    \hline
    CBCT&3D&Fixed center&Computed axial tomography&Orthodontics, tumor surgery \cr
    PX&2D&Moving center&Focal plane tomography&Tooth pulling/planting\cr
    \hline
    \end{tabular}
    \label{tab:ct_and_px}
\end{table*}

\section{Background and Related Works}

\subsection{Radiology in dental imaging}
There are mainly two radiological imaging methods in dental health, i.e., CBCT and PX. CBCT generates a 3D image of the oral cavity with rich spatial information of teeth, thus widely used in orthodontics and tumor surgery. As a comparison, PX is a faster and lightweight method used in the examination before pulling or planting teeth, where a 2D panoramic picture is taken of all the teeth along the mandibular curve. We show illustrations of these two imaging methods viewed in the axial plane in Fig.~ \ref{fig:imaging} and comparisons in Table \ref{tab:ct_and_px}. In CBCT, as shown in the left image, the X-ray tube and the film moves around a fixed center for $360^\circ$. The 3D image is then reconstructed from sinogram signals in 2D space \cite{ct_theory}, which is feasible as each point is projected from different directions during the imaging. In PX, the X-ray tube and the film move around a moving center from one side to the other. The trajectory, also named the focal plane, generally fits the curved shape of the mandible, leading to different projection levels for tissues at various locations. For example, as shown in the right picture, the red point at $O_1$ located on the moving trajectory is projected twice while the green point at $Q_1$ off the moving trajectory is projected only once when the X-ray tube and the film move from A to B. Therefore, the image shows stronger signals for tissues at $O_1$ than $Q_1$, thus generating a clear picture of objects around the focal plane. Like CBCT, points around the focal plane also receive multiple projections in PX but are not used to recover any 3D information during imaging. This feature is taken advantage of our proposed model for 3D oral reconstruction.

\subsection{Implicit representation in 3D reconstruction}
Implicit representation has been demonstrated to be a promising method in the task of 3D reconstruction since the work of neural radiance field (NeRF) \cite{nerf}, where the researchers use the deep neural network to map 5D coordinates of spatial location and viewing direction into the density and emitted radiance of a voxel. During the inference, the model could generate images from any position by rendering along the rays sent from the observation point. Based on this framework, D-NeRF \cite{dnerf} takes time as additional input to the system for the reconstruction of dynamic scenes. Nerfies \cite{nerfies} use an additional continuous volumetric deformation field to generate deformable photo-like scenes. Although our method also utilizes implicit 3D representation, there are still big differences due to the characteristic of the imaging process in radiology: 1) The movement of an X-ray tube has less degree of freedom (DoF) than a camera, thus leading to limited projection rays in both directions and origins. 2) The predicted density distribution represents the values of HU instead of the differential probability. 3) The reconstruction object should be view-independent.

\subsection{Cross-dimension translation in radiology}
Cross-dimension translation in radiology images between 2D and 3D by deep neural networks starts from the work of \cite{x_to_3d}, where the authors use an encoding-decoding network to learn a back projection function that maps a 2D projection image into 3D density volumes for the skull of mammals. Following this work, \cite{x2ct_gan, knee_3d} improve the reconstruction quality for the abdomen and knees by utilizing bi-planar X-ray images and adversarial networks. In dental healthcare, Oral-3D \cite{oral_3d} first uses a single panoramic X-ray image to reconstruct the 3D oral structure. X2Teeth \cite{x2_teeth} trains three networks to reconstruct and segment the teeth in 3D space with annotated X-ray images. Our model can be seen as an extension of these works that focus on the same problem but with a different technical solution: 1) In contrast to learning explicitly by auto encoding or adversarial learning, our method learns the representation of the 3D object in an implicit way. 2) Our model relies no more on paired 2D and 3D images or individual prior knowledge to restore the mandibular curve.

\section{Methodologies}

\begin{figure*}[t]
    \centering
    \includegraphics[width=\textwidth]{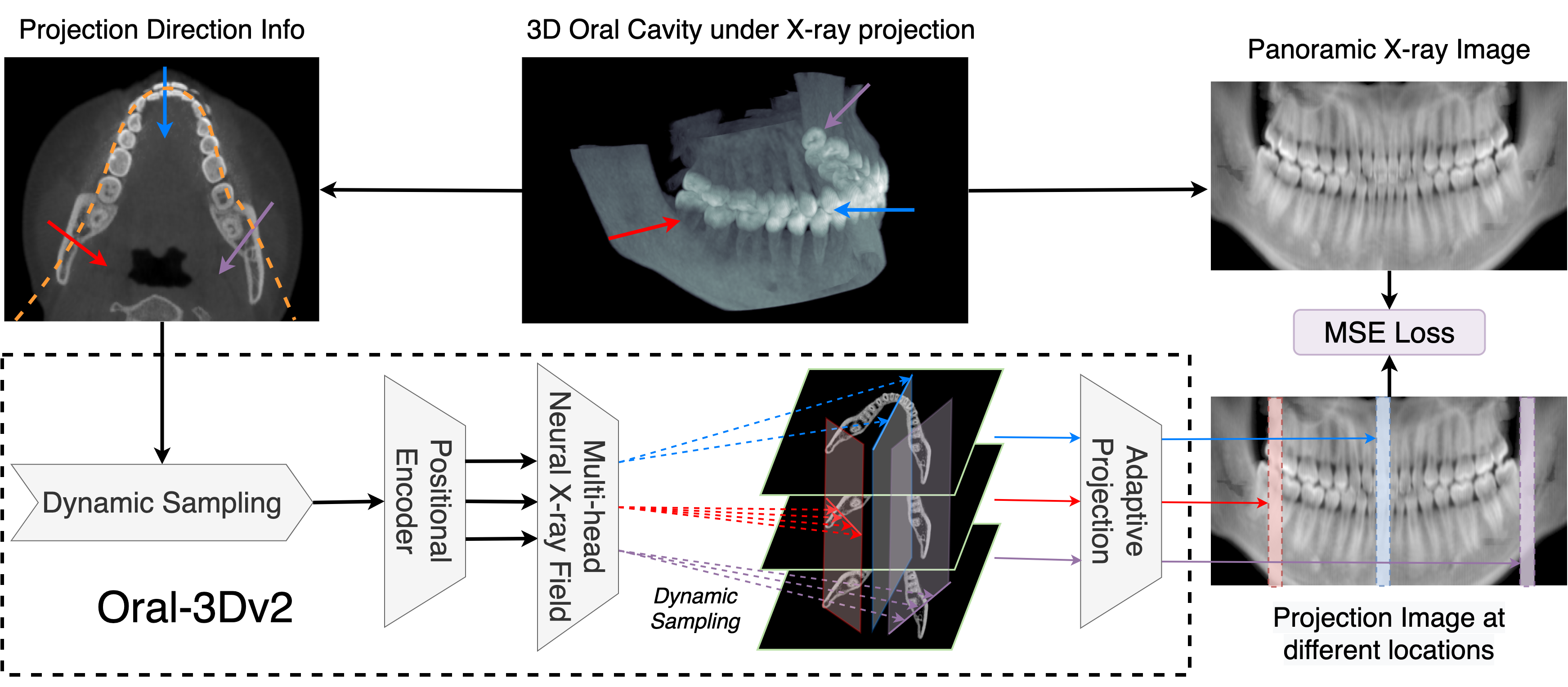}
    \caption{This image provides an overview of our model, i.e., Oral-3Dv2. Starting with radiation rays, we use a dynamic sampler to acquire sample points on each ray at random sampling rates. Then, we employ our proposed multi-head neural X-ray field (NeXF) with a positional encoder to predict densities in the 3D space. The NeXF outputs a bunch of HU values from a single 2D coordinate. Next, we generate a projection image adapting to the dynamic resolution during sampling. Finally, we calculate the MSE loss between the projection slice and the ground-truth image to update parameters of our implicit representation model.
    }
    \label{fig:model}
\end{figure*}

\subsection{Problem Definition}
Given a pair of projection image $I$ and the trajectory of the rotation center $O$ during the PX imaging, the object is to find an implicit 3D representation $V: \mathbf{p} \rightarrow h$ that maps 3D coordinates $\mathbf{p}$ into HU values $h$ and minimizes the mean square error against the projection image given the imaging function $F(\cdot)$. The problem can be defined as:
\begin{equation}
\label{eq:opt_overall}
    \mathop{\arg\min}\limits_{V(\cdot)} ||F(V, O) - I||_2.
\end{equation}
With the sampled rotation center point at $O_i$ and the corresponding projection image $I_i$, the reconstruction problem in Eq (\ref{eq:opt_overall}) could be solved by optimizing the below objective function:
\begin{equation}
\label{eq:opt_detail}
    L_{obj}=\sum_{i=1}^{N} ||f(V(\mathbf{p_1}), V(\mathbf{p_2}), \cdots, V(\mathbf{p_m})) - I_i||_2,
\end{equation}
where $\mathbf{p_1}, \cdots, \mathbf{p_m}$ are the coordinates of points sampled along the radiation ray sent from the X-ray tube, and $f$ is the projection function that maps multiple voxel values into a single one. To distinguish with existing NeRF-like models $V_{NeRF}$, we refer to our implicit representation model as $V_{NeXF}$ (short for neural X-ray field) to represent the field function in X-ray imaging.

\subsection{Overview}
We show an overview of our proposed model in Fig.~\ref{fig:model}, where paired rays and rendering results are taken as input to train the implicit representation model $V_{NeXF}$. Given the direction and origin of the projection ray inferred from the moving trajectory $T(O)$ of the X-ray tube, we first generate points along the radiation ray at a random sampling rate. The sampled coordinates are then taken as the input of a positional encoding module, followed by our proposed NeXF model, to generate the projection results. The model is updated according to Eq. \ref{eq:opt_detail} until converge. Although our framework looks similar to NeRF-like models, we have three major differences due to the feature of PX imaging, where the radiation rays are almost parallel to the axial plane. First, our NeXF has a multi-head structure, whose input is a 2D coordinate and output is a bunch of voxel values in the same axial location. Second, we use a dynamic sampling strategy instead of a pair of coarse and fine networks to improve the reconstruction quality. Third, our model is view-independent as it is unreasonable for various density values for the same voxel in radiology.

\subsection{Dynamic Sampling}
NeRF-based models generally utilize a pair of coarse and fine networks to determine the sampling rate along the rays due to multiple free spaces and occluded regions in their 3D objects viewed from the outside. However, this is not applicable to radiology as the aim of imaging is to observe the inside structure of the object. Therefore, points along the radiation rays should be evenly sampled to evenly indicate the density variance in 3D space. To accommodate this, we propose a dynamic sampling strategy that acquires points from radiation rays in a random resolution to improve spatial smoothness without introducing additional new networks. As shown in Fig. \ref{fig:model}, radiation rays sent from different directions (represented by the red, blue, and purple arrows) acquire different numbers of sampling points when generating projection images. We show that the variance in sampling rate during projection in training could significantly improve the reconstruction quality in the ablation experiments.

\subsection{Positional Encoding}
Positional encoding has been widely used in implicit 3D representation models due to the tendency of learning low-frequency details as revealed in recent research like \cite{positional_bias1, positional_bias2}. To solve this spectral bias problem, frequency encoding is introduced in \cite{nerf, freq_encoding1} to encourage the model to exploit high-dimension spatial information during reconstruction. We follow the same way as in \cite{attention} that utilizes multi-resolution sequence to encode the coordinate value $p$ from $\mathbf{p}$ into $L$ levels of embedding as: 

\begin{equation}
\begin{aligned}
{Enc}(p)= & (\sin \left(2^0 p\right), \sin \left(2^1 p\right), \cdots, \sin \left(2^{L-1} p\right. \\
& \left.\cos \left(2^0 p\right), \cos \left(2^1 p\right), \cdots, \cos \left(2^{L-1} p\right)\right)
\end{aligned}
\end{equation}

\begin{figure}[t]
    \centering
    \includegraphics[width=0.4\textwidth]{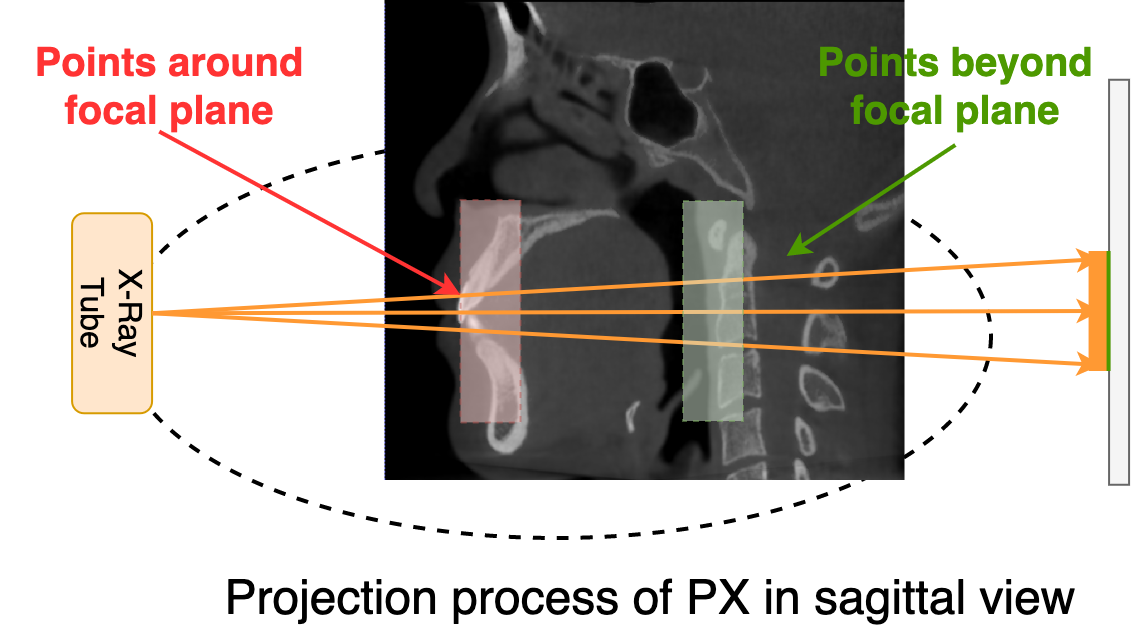}
    \caption{We show the imaging process of PX in sagittal view in this picture. Unlike cameras, the X-ray tube used in radiology has a fixed trajectory and direction when moving, thus leading to limited variance in direction of radiation rays.
    }
    \label{fig:px_sagittal}
\end{figure}

\subsection{Multi-head Neural X-ray Field}
\begin{figure}[t]
    \centering
    \includegraphics[width=0.45\textwidth]{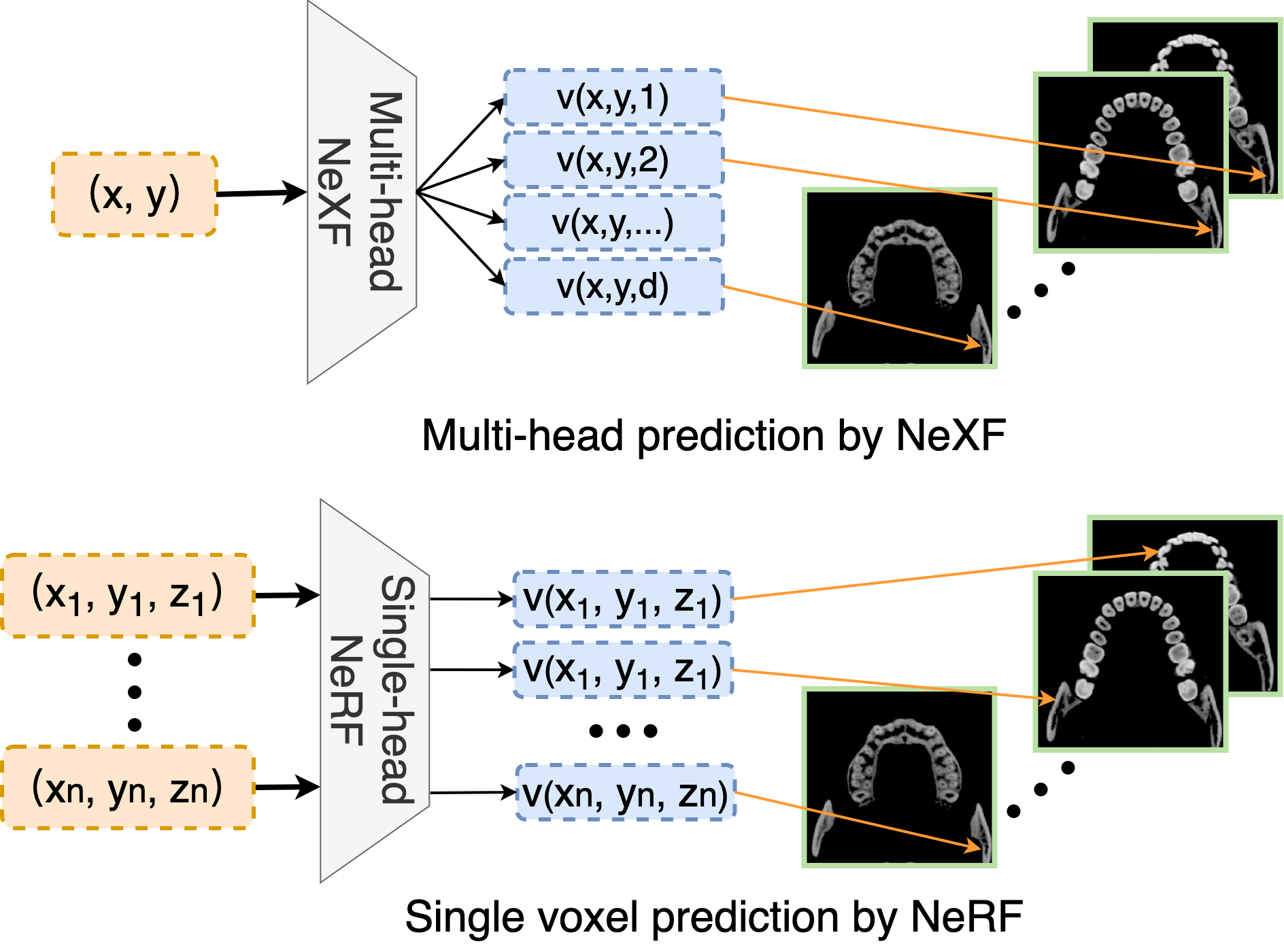}
    \caption{We show the comparison of implicit representation model between NeXF and NeRF models in this picture. The NeRF-like models have a single-head structure that outputs the specific voxel value of the given input. However, in NeXF the model only takes in a 2D coordinate by predicts a bunch of voxel values with its multi-head architecture. This architecture could best fit the imaging process of PX and significantly decrease the computation complexity during both training and inference.
    }
    \label{fig:multihed}
\end{figure}

Different from NeRF-based models, where the camera has more freedom in position and angle, the X-ray tube in radiological scans generally moves in a fixed trajectory, leading to a limited direction and origin of radiation rays during the imaging. For example, as shown in the sagittal view of PX imaging in Fig. \ref{fig:px_sagittal}, radiation rays that pass through the oral cavity are approximately parallel to the axial plane. Taking advantage of this feature, we propose a different radiance field model that predicts a bunch of voxel values in the 3D space from a 2D coordinate. Given that the 3D object in radiology should be view-independent, our implicit representation model $V_{NeXF}$ can be defined as:
\begin{equation}
    V_{NeXF}: (x, y) \rightarrow (v_{x, y, 1}, v_{x, y, 2}, \cdots, v_{x, y, z_n}),
\end{equation}
in comparison to $V_{NeRF}$ defined as:
\begin{equation}
    V_{NeRF}: (x, y, z, \theta, \phi) \rightarrow v_{x, y, z}.
\end{equation}
We compare the difference between $V_{NeXF}$ and $V_{NeRF}$ in Fig. \ref{fig:multihed}. $V_{NeXF}$ uses a multi-head architecture that takes in a 2D coordinate as input and outputs $z_n$ number of voxel values, where $z_n$ is the same as the resolution of reconstruction object in $z$ axis. In contrast, $V_{NeRF}$ only predicts a single value per 5D coordinate. Therefore, $V_{NeXF}$ can reduce the computational complexity from $\mathcal{O}(N^3)$ to $\mathcal{O}(N^2)$ compared with $V_{NeRF}$ during the reconstruction of $N \times N \times N$ object.

\begin{figure*}[t]
    \centering
    \includegraphics[width=0.9\textwidth]{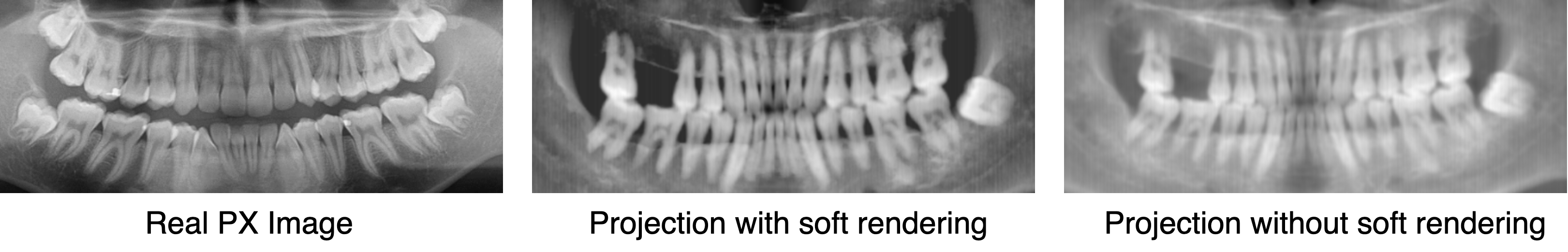}
    \caption{Comparison of different rendering methods in PX imaging. We can see that with soft rendering the generated PX image has a closer contrast with the real PX image (obtained from Internet). The real PX image looks more clear due to the high resolution of the PX machine.
    }
    \label{fig:px_compare}
\end{figure*}

\begin{table*}[ht]
    \centering
    \caption{Evaluation of 3D oral reconstruction by PSNR, SSIM(\%), and Dice.}
    \label{tab:compare}
    \begin{tabular}{p{6.0cm}<{\centering}p{2.0cm}<{\centering}p{2.0cm}<{\centering}p{2.0cm}<{\centering}p{2.0cm}<{\centering}}
    \hline
    Method&PSNR&Dice&SSIM&Overall\cr
    \hline
    NAF \cite{naf}&18.35$\pm$0.86&57.20$\pm$3.94&60.69$\pm$2.69&65.93\cr
    GAN \cite{gan}&16.71$\pm$0.89&75.10$\pm$1.46&63.96$\pm$7.03&76.93\cr
    ResEncoder \cite{x_to_3d}&18.26$\pm$0.50&72.67$\pm$1.56&62.52$\pm$5.56&75.49\cr
    Oral-3D \cite{oral_3d}&18.59$\pm$0.70&76.88$\pm$1.26&65.94$\pm$4.24&78.60\cr
    \textbf{Ours}&\textbf{20.34}$\pm$0.62&75.34$\pm$3.95&\textbf{81.06}$\pm$1.61&\textbf{86.04}\cr
    \hline
    \end{tabular}
\end{table*}

\subsection{Adaptive Projection}
Following Beer-Lambert absorption-only model \cite{beerlaw}, the fraction $\alpha$ of radiation arriving at the film after traveling volumes with spatially-varying density $\mu(t)$ along a ray parameterized with variable t within $[t_n, t_f]$ could be expressed as:
\begin{equation}
\label{eq:beer_law}
\alpha = \exp \int_{t_n}^{t_f}\mu(t)dt,
\end{equation}
where $\mu(t)$ is the attenuation coefficient and could be converted into a HU value by:
\begin{equation}
H(\mu)=1000 \times \frac{\mu-\mu_{\mathrm{water}}}{\mu_{\mathrm{water}}-\mu_{\mathrm{air}}},
\end{equation}
where $\mu_{\mathrm{water}}$ and $\mathrm{water}$ are constant values and $\mu$ is the accumulated attenuation coefficient along the ray path. By sampling along the radiation ray, Eq. (\ref{eq:beer_law}) can be converted into:
\begin{equation}
\label{eq:beer_law_discrete}
\alpha = \exp(\sum_{i}^{\lfloor t_f-t_n \rfloor}\mu_{i}).
\end{equation}
Therefore, the projection function $f(\cdot)$ in Eq. (\ref{eq:opt_detail}) can be adaptively expressed with our proposed implicit representation model $V$ and the dynamic sampling rate $N_s$ as:
\begin{equation}
\label{eq:render_function}
\begin{aligned}
f(\cdot) &= H(\frac{\sum_{i}^{\lfloor N_s(t_f-t_n) \rfloor}\mu_{i}}{N_s}) \\
&= H(\frac{\sum_{i}^{\lfloor N_s(t_f-t_n) \rfloor} H^{-1}(V(\mathbf{p_i})}{N_s})),
\end{aligned}
\end{equation}
where ${p_i}$ is the $i$-th sample point within $[t_n, t_f]$.

\section{Experiments}

\subsection{Dataset}
We collect a dataset consisting of 80 CBCT dental scans as groundtruth of the 3D oral structure and source images to simulate PX imaging. We divide the model into two groups: 1) 60 cases used for training models based on auto-encoding and adversarial learning, and 2) 20 cases used for inference and validation for all models. The CBCT scan is resized into a size of $288\times256\times160$ using trilinear interpolation to minimize influence brought by imaging machines.

\subsection{PX Imaging Simulation from CBCT}
The moving trajectory of rotation center in PX imaging is fitted by the beta function as:
\begin{equation}
    y = 256 - beta(x/288, 3.6, 3.6) * 100 - 25.
\end{equation}
We split the trajectory curve equally into 576 pieces and assume the radiation rays evenly cross each small curve in angles between $-\pi/4$ and $\pi/4$. Research in \cite{cbct_no_hu_1}\cite{cbct_no_hu_2} show that HU is unreliable in CBCT scans due to variations in gray-scale values for different areas in the scan, especially when the imaging areas have the same density but different relative positions. Therefore, we follow the same method proposed in \cite{PX_generation, oral_3d} during projection to get realistic PX images from CBCT. Then the projection function $f(\cdot)$ in Eq. (\ref{eq:render_function}) can be rewritten into $\hat{f}(\cdot)$:
\begin{equation}
\label{eq:soft_render_discrete}
    \hat{f}(\cdot)=S\cdot\log(\sum_i^{\left\lfloor N_s\left(t_f-t_n\right)\right\rfloor} e^{\frac{V(\mathbf{p_{i}})+C}{S}})-\log{N_s} - C,
\end{equation}
where $C=H(\mu_{air})$. Comparisons among real PX image and simulated images generated by $f(\cdot)$ and $\hat{f}(\cdot)$ can be seen in Fig \ref{fig:px_compare}, where PX images simulated by $\hat{f}(\cdot)$ has a more closer contrast as real images.

\subsection{Hyper-parameters and Network Architecture}
We select $S=1200$ in Equation (\ref{eq:soft_render_discrete}) to distinguish air and soft tissues in HU. The sampling rate $N_s$ for each radiation ray during training follows a uniform distribution in $[0.25, 1,25]$. The level $L$ used in positional encoding is selected to be 16 with the normalizaton of coordinates into $[-1, 1]$. For the multi-head NeXF, we use a 12-layer fully-connected neural network with residual connections and set the number of heads as 160, which is consistent with the CBCT data.

\subsection{Training and Evaluation}
The model is trained for 20k iterations with a batch size of 64. The model is optimized by Adam with a learning rate of 0.0001. We use structural similarity index measure (SSIM) \cite{ssim}, dice coefficient (DC), and peak signal-to-noise ratio (PSNR) to evaluate the reconstruction results. We also use the averaged score proposed in \cite{oral_3d} as the overall metric.

\subsection{Baseline Models}
We compare our method with baseline models that can be grouped into two categories. The first group including GAN \cite{gan}, Oral-3D \cite{oral_3d}, and Res-Encoder \cite{x_to_3d}. These models are trained with the 60 paired simulated X-ray images and CBCT images and learn the prediction of explicit 3D representation with either adversarial learning or auto-encoding. We put our model in the second group with NAF \cite{naf}, another implicit representation with the same framework attenuation coefficients in 3D space.
\section{Results}
\label{sec:results}

\begin{figure*}[t]
    \centering
    \includegraphics[width=\textwidth]{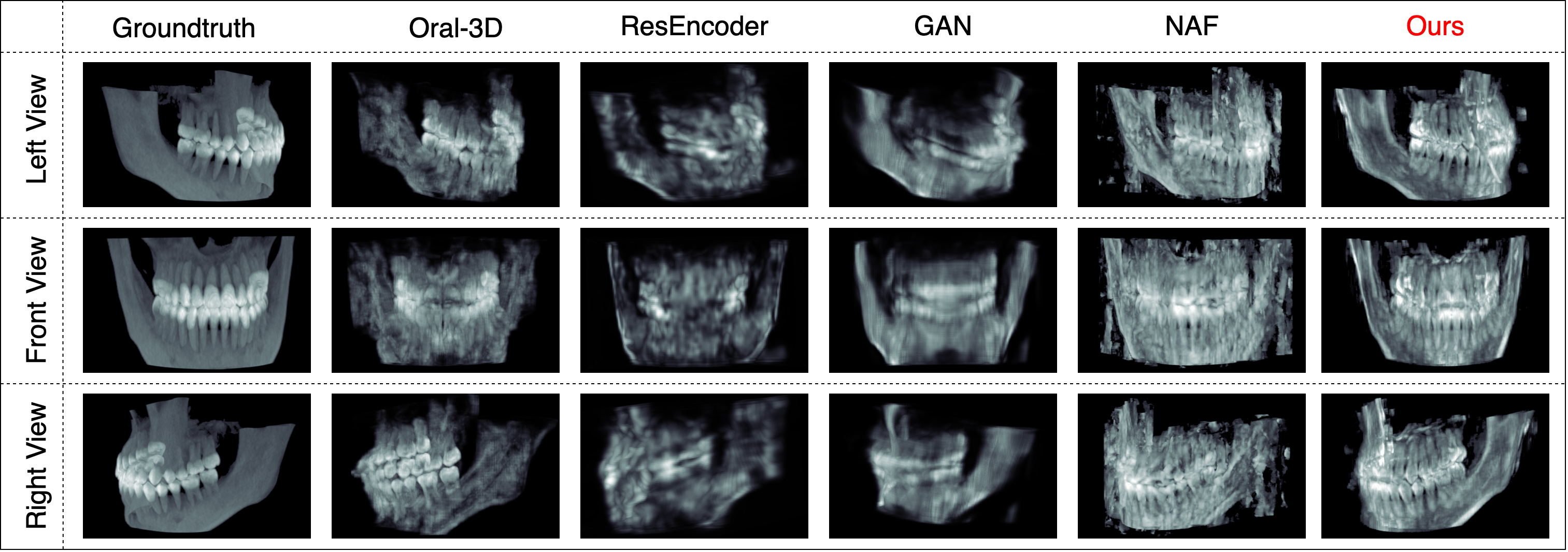}
    \caption{Comparison of 3D oral reconstruction by different methods from PX imaging. The reconstruction results are shown by maximum projection to compare density details. We could easily find that our method show the best performance with clear density density distributions and teeth boundaries.
    }
    \label{fig:3d_compare}
\end{figure*}

\begin{table*}[tp]
    \centering
    \caption{Ablation study by removing each component in our proposed method. M: Multi-head Prediction, D: Dynamic Sampling, P: Change $\hat{f}(\cdot)$ to $f(\cdot)$ in training}
    \label{tab:ablation}
    \begin{tabular}
     {p{0.8cm}<{\centering}p{0.8cm}<{\centering}p{0.8cm}<{\centering}p{2.2cm}<{\centering}p{2.2cm}<{\centering}p{2.2cm}<{\centering}p{1.5cm}<{\centering}p{1.5cm}<{\centering}}
    \hline
    M&D&P
    &PSNR&SSIM(\%)&Dice(\%)&Overall&Drop\cr
    \hline
    \xmark&\cmark&\cmark&16.68$\pm$0.74&73.62$\pm$5.49&61.25$\pm$4.57&72.76&-13.28\cr
    \cmark&\xmark&\cmark&16.80$\pm$0.71&61.44$\pm$5.87&73.29$\pm$3.24&72.91&-13.13\cr
    \cmark&\cmark&\xmark&16.57$\pm$1.08&63.63$\pm$3.07&70.28$\pm$3.28&72.25&-13.79\cr
    \hline
    \end{tabular}
\end{table*}

\subsection{Qualitative Comparison}
We first show qualitative comparison in Fig.~\ref{fig:3d_compare} to compare the reconstruction results of baseline models. We can see that although ResEncoder and GAN could restore the curved shape of mandible without any prior knowledge, these models fail to recover the detail density distribution in the reconstruction results. For NAF, the model could recover the curved shape and density variance. But the results contain too much noise and is hard to identify the teeth shape. For Oral-3D, the model could restore both shape and teeth details with the help of individual dental arch shape. However, its reconstruction quality is obviously lower than our method, especially for the details of density change between teeth root and the mandible.

\subsection{Quantitative Comparison}
We then show the quantitative comparison by the proposed metrics in Table \ref{tab:compare}. The dice score is computed by setting a threshold at 500 HU to extract the bone from soft tissues. We could see our model could significantly outperform other models, with improvement of $+5$ in SSIM and $+7.5$ in the overall score against the state-of-the-art method without training on paired images or deformation by individual prior knowledge. To be noted, Oral-3D has a better Dice score but lower performance in PSNR and SSIM. This is consistent with the visualized results shown in Fig.~\ref{fig:3d_compare}, where Oral-3D restores less density details than our model.

\subsection{Ablation Study}
\label{sec:ablation}
We conduct an ablation study to evaluate the contribution of each component in our model: 1) replace the multi-head field function with a single-head predictor and taking in 3D coordinates as input for the positional encoder; 2) use a fixed sampling rate of $N_s=1$ to generate sample points on projection rays; 3) change the rendering function in Eq. (\ref{eq:soft_render_discrete}) to Eq. (\ref{eq:render_function}) when training the model. We use the letters M, D, and P to represent these changes. Results are shown in Table~\ref{tab:ablation}, where the performance drops significantly (about -13 in Overall) when changing any module. We could see the dynamic sampling strategy can greatly improve the reconstruction quality without introducing additional models. And the multi-head architecture has stronger ability in implicit representation in radiation imaging. 

\subsection{Complexity Analysis}
We also conduct the complexity analysis for NeXF and NeRF with the result in Table \ref{tab:complexity}, where the models run on a Nvidia A100 GPU with the same batch size. Combined with experiments in the ablation study, we see that NeXF is both efficient and effective over NeRF, i.e., $\times3.3$ faster in training and inference. To be noted, the training process only takes 30 minutes, which is quite acceptable in clinical application compared to a CBCT scan that takes about 20 minutes \cite{cbct_time}.
\begin{table}[tp]
    \centering
    \setlength\tabcolsep{2pt}
    \caption{Complexity analysis between NeXF and NeRF.}
    \begin{tabular}{p{1.6cm}<{\centering}p{1.8cm}<{\centering}p{1.8cm}<{\centering}p{1.8cm}<{\centering}}
    \hline
    Method&Train(ms/iter)&Infer(s)&Learn(min)\cr
    \hline
    NeXF &90&20&30\cr
    NeRF &300&60&100\cr
    \hline
    \end{tabular}
    
    \label{tab:complexity}
\end{table}

\section{Conclusion}
In this paper, we propose a new method for reconstructing the 3D oral structure from projection information in panoramic X-ray imaging. We utilize an implicit representation model with multi-head architecture to accommodate the imaging process of PX and a dynamic sampling strategy to refine the reconstruction results. Unlike existing deep learning models like Oral-3D, our method does not require extensive patient data or dense annotations to reconstruct the complicated structure of oral cavity. Extensive experiments show that our model significantly outperforms state-of-the-art models both qualitatively and quantitatively with clear density details of teeth and the mandible in the reconstructed oral structure. Furthermore, the complexity analysis show that our method has great potential in clinical applications with the low radiation and comparable reconstruction speed.

\clearpage
\bibliography{aaai24}

\end{document}